\documentclass[journal]{IEEEtran}

\usepackage{graphicx}
\usepackage{cite}
\usepackage{xfrac}
\usepackage{amsmath}
\usepackage{mathtools}
\usepackage{xcolor}

\begin{document}

\title{Automated Optic Nerve Head Detection Based on Different Retinal Vasculature Segmentation Methods and Mathematical Morphology}
%
%
\author{Meysam Tavakoli,
        Mahdieh Nazar,
        Alireza Golestaneh,
        Faraz Kalantari
\thanks{M. Tavakoli is with the Dept. of Physics, Indiana University-Purdue University, Indianapolis, IN, USA.}
\thanks{A. Golestaneh is with the Electrical Engineering Department, Arizona State University, Tempe, AZ, USA}
\thanks{M. Nazar is with the Biomedical Sciences Department, Shahid Beheshti Medical Sciences, Tehran, IRAN}
\thanks{F. Kalantari is with Department of Radiation Oncology, University of Texas Southwestern Medical Center, Dallas, TX, USA 
}
\\
\textbf{To appear in: 2017 IEEE Nuclear Science Symposium and Medical Imaging Conference (NSS/MIC)\\
DOI: 10.1109/NSSMIC.2017.8532764}
}

\maketitle

\begin{abstract}
Computer vision and image processing techniques provide important assistance to physicians  and relieve their work load in different tasks.
In particular, identifying objects of interest such as lesions and anatomical structures from the image is a challenging   and iterative process that can be done by using computer vision and image processing approaches in a successful manner. 
Optic Nerve Head (ONH) detection is a crucial step in retinal image analysis algorithms.
The goal of ONH detection is to find and detect other retinal landmarks and lesions  and their corresponding diameters, to use as a length reference to measure objects in the retina.
The objective of this study is to apply three retinal vessel segmentation methods, Laplacian-of-Gaussian edge detector, Canny edge detector, and Matched filter edge detector for detection of the ONH either in the normal fundus images or in the presence of retinal lesions (e.g. diabetic retinopathy).  
The steps for the segmentation are as following: 1) Smoothing: suppress as much noise as possible, without destroying the true edges, 2) Enhancement: apply a filter to enhance the quality of the edges in the image (sharpening), 3) Detection: determine which edge pixels should be discarded as noise and which should be retained by thresholding the edge strength and edge size, 4) Localization: determine the exact location of an edge by edge thinning or linking.
To evaluate the accuracy of our proposed method, we compare the output of our proposed method with the ground truth data that collected by ophthalmologists on retinal images belonging to a test set of 120 images.
As shown in the results section, by using the Laplacian-of-Gaussian vessel segmentation, our automated algorithm finds 18 ONHs in true location for 20 color images in the CHASE-DB database and all images in the DRIVE database. For the Canny vessel segmentation, our automated algorithm finds 16 ONHs in true location for 20 images in the CHASE-DB database and 32 out of 40 images in the DRIVE database. And lastly, using matched filter in the vessel segmentation, our algorithm finds 19 ONHs in true location for 20 images in CHASE-DB database and all images in the DRIVE.

\end{abstract}

\begin{IEEEkeywords}
Diabetic retinopathy, image processing, Optic Nerve Head, retinal blood vessel, Canny edge detector, Laplacian-of-Gaussian edge detector, Match filter
\end{IEEEkeywords}

%
\IEEEpeerreviewmaketitle

\section{Introduction}
\begin{figure}[h!]
\includegraphics[width=\linewidth]{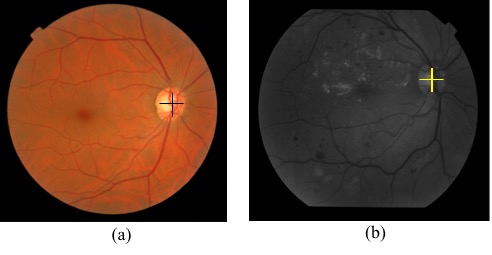}
\centering
\caption{Fundus image from MUMS-DB (a) normal color image (b) green channel.}
\label{fig0}
\end{figure}

\IEEEPARstart{T}{he} The computer techniques are applied for providing physicians assistance at any time and to relieve their work load or iterative
works as well, to identify the object of interest such as lesions and anatomical structures from the image \cite{TavakoliFA, Mehdizadeh-color, Tavakoli-SPECT}.
The identification of the optic nerve head (ONH) or optic disk (OD) is important in retinal image analysis, to locate anatomical
components in fundus images, for vessel tracking, as a reference length for measuring distances in retinal images, and for
registering changes within the ONH region because of some diseases Diabetic retinopathy (DR) or glaucoma. 
ONH is yellowish region in color retinal image that
usually covered one seventh of fundus image \cite{sinthanayothin, Tavakoli-ONH1}. The main characteristic of ONH is rapid intensity changing due to dark thick blood
vessels that are in vicinity of bright ONH. This intensity flactuation is the characteristic of interest for ONH recognition. In the
other words, ONH is usually the brightest component on the fundus, and therefore a collection of high intensity pixels with a high
grey-scale value will identify the location of ONH \cite{tavakoli-twopreprocessingsteps, Morales S} (See Fig.~.\ref{fig0}).
ONH have three properties in other to be detected: (1) ONH appears as a bright disk nearly 1600$\mu$m  in diameter (2) the large blood vessels enters from its above and blew (3) blood vessel diverge from ONH \cite{sinthanayothin, Muangnak N, Constante P}.
Some advantages of detecting ONH are: (1) the identification of ONH is critical for automatic detection of some anatomical
structures and retinal lesions. One of the elements for extraction in this situation is vessel tree especially large vessels that are
located in adjacent of the ONH \cite{Abdullah M, Sherwani SM}. (2) In other hand, detecting and masking ONH can decrease false positive rate of lesions
that detected in identification of some diseases like DR \cite{Gagnon L, Sengar}.


 




\section{Previous Work}
There are several algorithms that detect the location of ONH, center of that or its boundary \cite{Abdullah M, Tavakoli-ONH1, Abdel-Ghafar, OsarehONH1, ForacchiaWorkshop, Hoover1, foracchia, Walter, hoover-fuzzyconvergence, Tavakoli-ONH2, youssif, aquino, li, Aquino, Chrastek}. However, identification of ONH is difficult because of discontinuity of its
boundary, due to crossing large vessels as well as considerable color or intensity conversion in some parts of retinal image
because of some pathologies (such as exudates in color image). The study introduced effective approach based on active
contour model has been reported by Osareh et al. that was complex and time consuming. Firstly, the image was normalized by
using histogram specification, and then the ONH region was averaged from 25 color-normalized images, to determine a gray-
level template. Then the normalized correlation coefficient was applied to find the finest match between the template and all the
candidate pixels in the given image \cite{OsarehONH1}. Another appropriate technique is belonging to Abdel-ghafar et al \cite{Abdel-Ghafar}. In this study they
used the circular Hough transform for detecting the ONH which has a roughly circular shape, Hough transform make it possible
to find geometric shapes within an image. The retinal vessel network in the green-channel image was suppressed by using the
closing morphological operator. For extracting the edges in the image the Sobel operator and a simple threshold were then
used. Finally circular Hough transform was employed to the edge points, and the largest circle was determined consistently to
correspond to the ONH. All of these studies detected ONH based on its shape and color. On the other hand, some algorithms recognized ONH according to tracking vessels until their origin \cite{ForacchiaWorkshop, Hoover1}. Also, Foracchia et al. have been reported on a new
technique for detecting the ONH using a geometrical parametric model (retinal vessels originating from the ONH and their path
follows a similar directional pattern in all images) to describe the typical direction of retinal vessels as they converge on the
ONH \cite{foracchia}. Sinthanayothin et al. identified the location of the ONH employing the variance of intensity between the ONH and
adjacent blood vessels \cite{sinthanayothin}. At first they preprocessed by using an adaptive local contrast enhancement method that was used to
the intensity component. Instead of applying the average variance in intensity and assuming that the bright appearing
retinopathies (e.g., exudates) are far from the ONH size, Walter and Klein \cite{Walter}, estimated the ONH center as the center of the
biggest brightest connected object in a fundus image. They obtained a binary image that consisted all the bright regions by
thresholding the intensity image. Moreover, Hoover and Goldbaum correctly identify ONH location by using a ‘‘fuzzy
convergence’’ algorithm (finds the strongest vessel network convergence as the primary feature for detection using blood
vessel binary segmentation, the disc being located at the point of vessel convergence. Brightness of the optic disc was used as
a secondary feature) \cite{hoover-fuzzyconvergence}. Poureza and Tavakoli \cite{Tavakoli-ONH1} located the ONH using Radon transform combine with some overlapping sliding window. At first, they chose the blue channel of the color retinal image and using the proposed method tried to detect the location of ONH. In another study, Tavakoli, et al. tried to detect the ONH using fluorescein angiography retinal images \cite{Tavakoli-ONH2}. They first used preprocessing method and after that using Radon transform locat the center of the ONH. \\
The objective of this study is to apply three retinal vessel segmentation methods, 1) Laplacian-of-Gaussian edge detector (using second-order spatial differentiation), 2) Canny edge detector (estimating the gradient intensity), and 3) Matched filter edge detector for detection of ONH either in normal fundus images or in presence of retinal lesion like in DR. 

\section{Proposed Method}

\begin{figure}[h!]
\includegraphics[width=\linewidth]{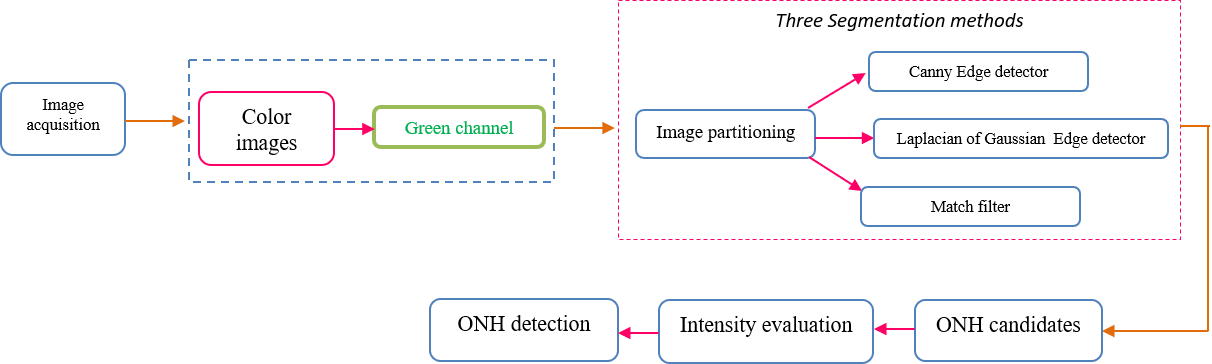}
\centering
\caption{The overall scheme of the methods for detection of the ONH}
\label{fig1}
\end{figure}

The overall scheme of the methods used in this study is shown in Fig.\ref{fig1}.


\subsection{Materials}

To detect the ONH, three databases (one rural and two publicly available databases) were used. 
The first rural database was named  Mashhad University Medical Science Database (MUMS-DB). 
The MUMS-DB provided 100 retinal images including 80 images with DR and 20 without DR. 
The fundus images were captured via a TOPCON (TRC-50EX) retinal camera at 50° field of view (FOV) and mostly obtained from the posterior pole view including ONH and macula with of resolution $2896\times1944$ pixels \cite{tavakoli-fluoresceinangiography, tavakoli-radon}. 
The second dataset was the DRIVE database consisting of 40 images with image resolution of $768\times584$ pixels in which 33 cases did not have any sign of DR and 7 ones showed signs of early or mild DR with a 45° FOV. This database is divided into two sets: testing and training set, each of them containing 20 images \cite{researchsection}. The last database, CHASE\_DB1 dataset includes 28 retinal images with image resolution of $999\times960$ pixels, acquired from both the left and right eye \cite{Fraz-CHASE1, Fraz-CHASE2, Fraz-CHASE3}.

\subsection{Preprocessing and Image Enhancement}

 The preprocessing step provides us with an image with high possible vessel and background contrast and also unifies the histogram of the images. 
 Although retinal images have three components (R, G, B), their green channel has the best contrast between vessel and background; so the green channel is selected as input image (I).  
First,  we used mathematical morphology operators. Mathematical morphology has been widely used in image processing and pattern recognition. Morphological operations work with two parts. The first one is the image to be processed and the second is called structure element. 
Erosion is used to reduce the objects in the image with the structure element, also known as the kernel. 
In contrast, dilation is used to increase the objects in the image. Secondary operations that depend on erosion and dilation are opening and closing operations. Opening, denoted as $f \circ b$, is applying an erosion followed by a dilation operation. The b represents the structure element. On the other hand, closing is first applying dilation then erosion. It is denoted as $f \bullet b$. Building from opening and closing operations, the top-hat transform is defined as the difference between the input image and its opening or closing.
The top-hat transform is one of the important morphological operators. 
Based on dilation and erosion, opening and closing denoted by $f \circ b$ and $f \bullet b$ are defined. 
The top-hat transform is defined as the difference between the input image and its opening. 
The top-hat transform includes white top-hat transform (WTH) and black top-hat transform (BTH) are defined by:

\begin{equation} \label{eq:sensitivity}
\begin{cases}
WTH(x,y)= f(x,y) - f \circ b(x,y) \\
BTH(x,y)= f \bullet b(x,y) - f(x,y) \\
\end{cases}
\end{equation}

\begin{figure}[h!]
\includegraphics[width=\linewidth]{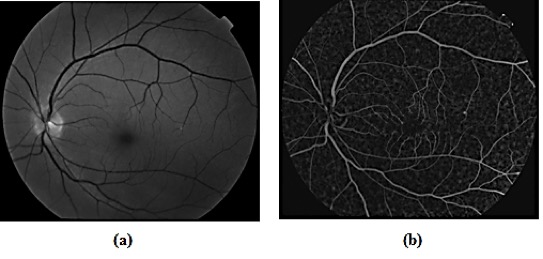}
\centering
\caption{(a) Fundus image from (b) top-hat result and contrast stretching (c) result of subtraction of top-hat and filtered top-hat image.}
\label{fig2}
\end{figure}

In our pre-processing the basic idea is increasing the contrast between the vessels and background regions of the image. WTH or BTH extract bright and dim image regions corresponding to the used structure element. 
Using the concept of WTH or BTH is one way of image enhancement through contrast enlarging based on top-hat transform.
In the fundus images, the background brightness is not the same in the whole image. This background variation would lead to missed vessels or false vessel detection in the following steps. Moreover, in I, background is brighter than the details, however the vessels and other components are preferred to appear brighter than background. To deal with the problem, I is inverted as shown in I= 255 - I.
Since we need a uniform background, to decrease the intensity variations in vessels background, we were firstly applied WTH(x,y) on image. It gave a high degree of differentiation between these features and background. A top-hat transformation was based on a ‘‘disk structure element’’ whose diameter was empirically found that the best compromise between the features and background. The disk diameter depended on the input image resolution.
After top-hat transformation, we used contrast stretching to change the contrast or brightness of an image. The result was a linear mapping of a subset of pixel values to the entire range of grays, from the black to the white, producing an image with much higher contrast \cite{tavakoli-radon}. The result of first step is shown in Fig.~\ref{fig2} 

\subsection{Procedure of ONH detection}
ONH can be explained as bright oval object placed versus a darker background with ill-defined boarder because of thick vessel come out and go in through it. 
When the image quality will be improved using preprocessed the candidate region of the ONH is obtained, the position of the ONH is identified using our model. Our approach  addresses the image locally and regionally where homogeneity of the ONH is more likely to happen. The algorithm is composed of 3 steps: Generation of sub-images, vessel segmentation, and ONH detection. 
In order to extract ONH, it should be extracted in local windows.
\subsubsection{Multi-overlapping window}
\begin{figure}[h!]
\includegraphics[width=2in]{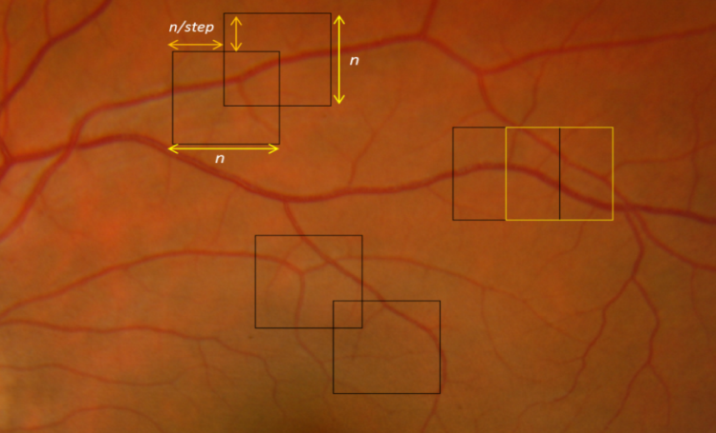}
\centering
\caption{Simple example of window size and overlapping ratio.}
\label{fig33}
\end{figure}

In the proposed algorithm, each fundus image was partitioned into overlapping widows in the first step. To find objects on
border of sub-images, an overlapping pattern of sliding windows was defined. For determining the size of each sub-image or
sliding window our knowledge database was used. In this regard, minimum and maximum sizes of targeted object specify the
size of the windows (n).
The window size (n) has a direct effect on the extraction accuracy. Another important parameter which affects the algorithm's accuracy is the windows overlapping \cite{Tavakoli-ONH1}. In Fig. \ref{fig33} we have shown some sample sub-images in a retinal image.

\subsubsection{Vessel Segmentation}
Here we applied three retinal vessel segmentation methods, 1) Laplacian of Gaussian edge detector \cite{Lowell J}, 2) Canny edge detector \cite{Canny, Zana}, and 3) Matched filter edge detector \cite{Chaudhuri} for detection of ONH either in normal fundus images or in presence of retinal lesion like in diabetic retinopathy (DR). 
In general, the steps for the edge detection are in following: (1) Smoothing: suppress as much noise as possible, without destroying the true edges, (2) Enhancement: apply a filter to enhance the quality of the edges in the image (sharpening), (3) Detection: determine which edge pixels should be discarded as noise and which should be retained by thresholding the edge strength and edge size, (4)	Localization: determine the exact location of an edge by edge thinning or linking.
In the Laplacian-of-Gaussian (LoG) edge detector uses the second-order spatial differentiation.
\begin{equation}
\bigtriangledown ^2 f = \dfrac{\partial^2 f}{\partial x^2} + \dfrac{\partial^2 f}{\partial y^2}
\end{equation}
The Laplacian is usually combined with smoothing as a precursor to finding edges via zero-crossings. The 2-D Gaussian function:
\begin{equation}
h(x,y) = e^{\dfrac{-(x^2 + y^2)}{2 \sigma ^2}}
\end{equation}
Where $\sigma$ is the standard deviation, blurs the image with the degree of blurring being determined by the value of $\sigma$. If an image is pre-smoothed by a Gaussian filter, then we have the LoG operation that is defined:
\begin{equation}
(\bigtriangledown ^2 G_{\sigma})*\textit{I}
\end{equation}
Where $\bigtriangledown ^2 G_{\sigma}(x,y)= (\dfrac{1}{2\pi \sigma^4})(\dfrac{x^2 + y^2}{2 \sigma ^2} - 2)e^{\dfrac{-(x^2 + y^2)}{2 \sigma ^2}}$ \\
In Canny edge detection, we estimate the gradient magnitude, and use this estimate to determine the edge positions and directions. 
\begin{equation}
\begin{cases}
f_{x} = \dfrac{\partial f}{\partial x} = K_{\bigtriangledown_{x}} * (G_{\sigma}*\textit{I}) = (\bigtriangledown_{x}G_{\sigma})*\textit{I}\\
f_{y} = \dfrac{\partial f}{\partial y} = K_{\bigtriangledown_{y}} * (G_{\sigma}*\textit{I}) = (\bigtriangledown_{y}G_{\sigma})*\textit{I}
\end{cases}
\end{equation}
Where 
\begin{equation}
\begin{cases}
\bigtriangledown_{x}G_{\sigma}=(\dfrac{-x}{2\pi \sigma^4})e^{\dfrac{-(x^2 + y^2)}{2 \sigma ^2}} \\\bigtriangledown_{y}G_{\sigma}=(\dfrac{-y}{2\pi \sigma^4})e^{\dfrac{-(x^2 + y^2)}{2 \sigma ^2}}
\end{cases}
\end{equation}

The algorithm runs in 4 separate steps: (1) Smooth image with a Gaussian: optimizes the trade-off between noise filtering and edge localization, (2) Compute the Gradient magnitude using approximations of partial derivatives, (3) Thin edges by applying non-maxima suppression to the gradient magnitude, and (4) Detect edges by double thresholding. We can compute the magnitude and orientation of the gradient for each pixel based two filtered images. \\
$|\bigtriangledown f(x,y)|= \surd f_{x}^2 + f_{y}^2$ = rate of change of f(x,y)\\
$\angle \bigtriangledown f(x,y) = tan^-1(\dfrac{f_{y}}{f_{x}})$= orientqtion of rate of f(x,y)

The matched filter has been widely used in the detection of blood vessels of the human retina digital image. In this paper, the matched filter response to the detection of blood vessels is increased by proposing better filter parameters.
The Matched Filter was first proposed in to detect vessels in retinal images. It makes use of the prior knowledge that the cross-section of the vessels can be approximated by a Gaussian function. Therefore, a Gaussian-shaped filter can be used to “match” the vessels for detection. The Matched Filter is defined as
\begin{equation}
G(x,y)=(\dfrac{1}{\surd 2\pi \sigma^2})e^{\dfrac{-(x^2 + y^2)}{2 \sigma ^2}} - m_{0}
\end{equation}
Where $m_{0}$ is chosen to make kernel G(x,y) has zero mean.\\

\subsubsection{ONH detection}
To detect the ONH we compare all sub-images with maximum density of vessels. To do this we compare all sub-images which have a peak profile, higher than a predefined threshold. In better word, we are looking for the ONH among those candidates which have maximum entropy of vessels. In fact when we segment the vessel in the next step we are looking for maximum entropy of the thick vessels. So we try to check the maximum entropy in each overlapping window. When we select our candidates the last step is comparing them using the intensity variation. In this case we pick that candidate with maximum intensity. \\
In the next step, to pick correct candidate which has the ONH, we look for the sub-images with highest intensity. Because as we mentioned in the introduction, ONH appears as a bright disk nearly 1600$\mu$m  in diameter in the retinal images.

\begin{figure}[h!]
\includegraphics[width=\linewidth]{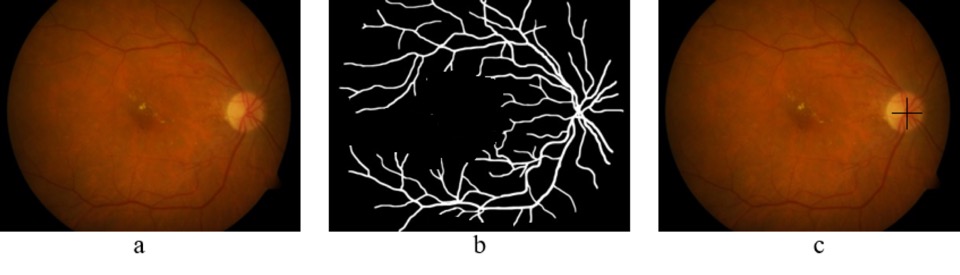}
\centering
\caption{(a) Input image from MUMS-DB; (b) result of vessel segmentation using Match filter; (c) final ONH detection.}
\label{fig3}
\end{figure}

\section{Experimental Results}

To calculate the efficiency of the current methods in ONH detection and also to compare the results with other reported studies, it is necessary to compare all pixels of the final automated segmentation  images with the manual segmentation or gold standard (GS) files. For the evaluation, we used the concept of sensitivity (Se) and specificity (Sp).
The results for the automated method compared to the GS were calculated for each image. The higher the sensitivity and specificity values, the better the procedure. These metrics are defined as: 

\begin{equation} \label{eq:sensitivity}
\begin{cases}
Sensitivity = \frac{TP}{TP+FN} \\
Specificity = \frac{TN}{TN+FP} \\
\end{cases}
\end{equation}

Where TP is true positive, TN is true negative, FP is false positive and FN is false negative.

\subsection{Training and Test Set for the Image Database}
\begin{figure}[h!]
\includegraphics[width=\linewidth]{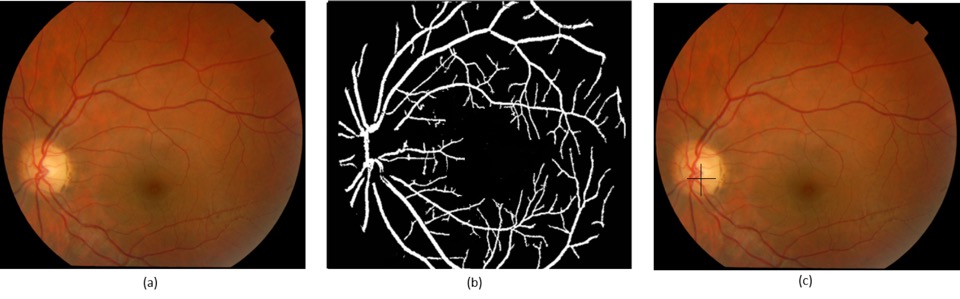}
\centering
\caption{(a) Input image from MUMS-DB; (b) result of vessel segmentation using Match filter; (c) final ONH detection.}
\label{fig4}
\end{figure}

\begin{figure}[h!]
\includegraphics[width=\linewidth]{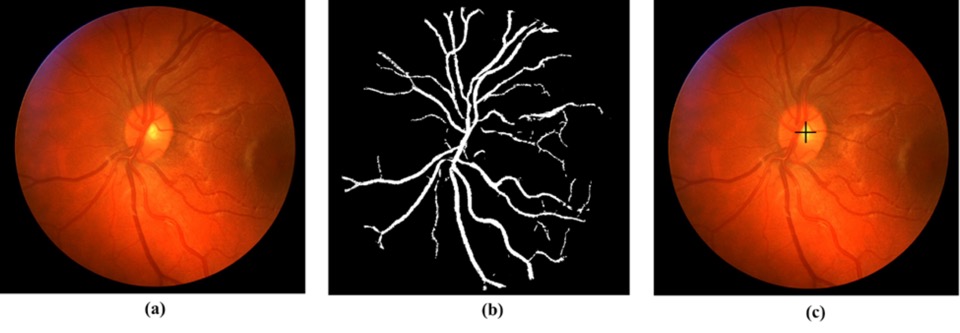}
\centering
\caption{(a) Input image from CHASE-DB; (b) result of vessel segmentation using Match filter; (c) final ONH detection.}
\label{fig5}
\end{figure}

In this study, we used 48 images for a training set (learning purpose). This consisted of 20 images from MUMS-DB, and DRIVE, and 8 images from CHASE-DB Databases. The test set (test purposes) consisted of 120 fundus images of which 80, 20, and 20 images from MUMS-DB, DRIVE, and CHASE-DB respectively. After fixing the parameters of our algorithm using training set, our algorithm was tested in each image of the databases in test set.  Some results are shown in Fig. \ref{fig3},  Fig. \ref{fig4}, and Fig. \ref{fig5} from MUMS-DB and CHASE-DB.

\subsection{Comparing the Statistics Results of ONH detetction in Three Databases}
The sensitivity of the threshold was also characterized along the Equation (8). A ROC curve is a plot o (Se) versus (1-Sp). 
A ROC curve, plotted to show the effect of a varying threshold, shows the presence or absence of sub-vessels in each sub-image, denoted by the $Th$ parameter, in our datasets. Parameters used for plotting ROC are shown in Table \ref{table2}.

\begin{table}[h!]
\renewcommand{\arraystretch}{1.3}
\caption{NUMBER OF PARAMETERS IN OUR ALGORITHM IN VESSEL SEGMENTATION FOR THE THREE DATABASES}
\label{table2}
\centering
\resizebox{\columnwidth}{!} {
\begin{tabular}{|c|c|c|c|c|c|}
\hline
Database & No. of Images & Window Size (n) & Step & Th \\
\hline\hline
MUMS-DB & 100 & 62 & 5 & [0,5] \\
\hline
DRIVE & 40 & 15 & 6 & [0,5] \\
\hline 
CHASE\_DB1 & 40 & 30 & 5 & [0,5] \\
\hline
\end{tabular}
}
\end{table}

\begin{figure}[h!]
\includegraphics[width=\linewidth]{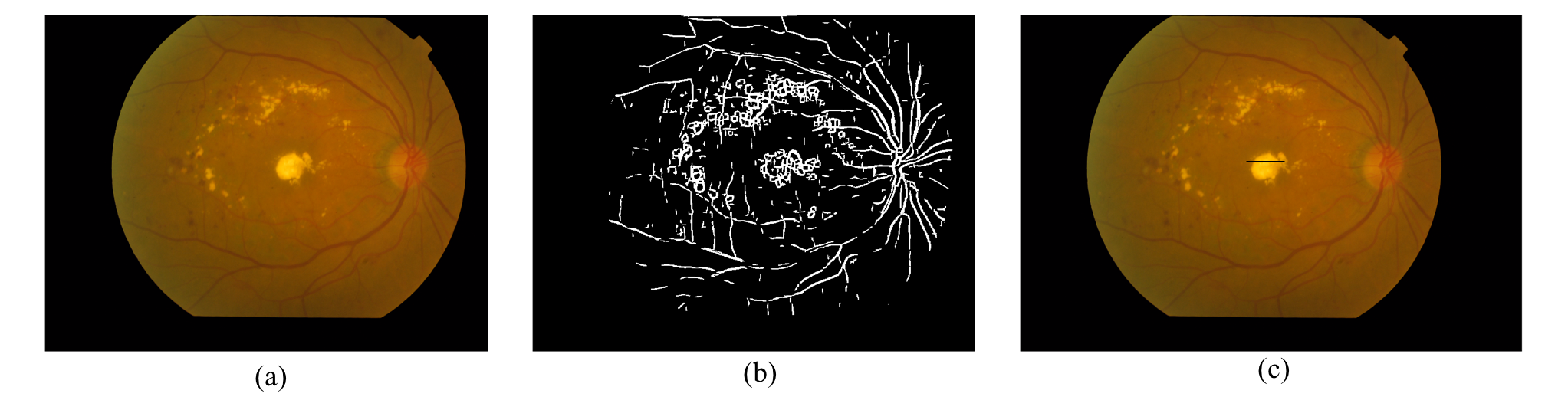}
\centering
\caption{(a) Input image from MUMS-DB; (b) result of vessel segmentation using Match filter; (c) final ONH detection Which is not correct).}
\label{fig8}
\end{figure}

Statistical information about the sensitivity and specificity measures is extracted. The higher the sensitivity and specificity values, the better the procedure. For all retinal images of test set (120 images), our reader labeled the ONH on the images and the result of this manual detections were saved to be analyzed further. According to manual ONH detection using the Laplacian-of-Gaussian vessel segmentation our automated algorithm finds 90\% of the ONHs (18 of ONH in true location for 20 color images) in CHASE-DB database and all images in DRIVE database (100\%). For the local database, MUMS-DB, the method detected the ONH correctly in 90\% of the ONH (72 images out of 80 images).
The Canny vessel segmentation our automated algorithm finds 15 of ONH in true location for 20 color images in CHASE-DB database (75\%) and 16 out of 20 images in DRIVE database (80\%). For the local database, MUMS-DB, our method detected the ONH correctly in 70 images out of 80 images (87.5\%).
At last, using Matched filter in the vessel segmentation our algorithm found the ONH with accuracy of 95\% (19 of ONH in true location for 20 color images) in CHASE-DB database and all images in DRIVE database (100\%). For the local database, MUMS-DB, the method detected the ONH correctly in 93.75\% of all fundus images (75 images out of 80 images). \\
\begin{figure}[h!]
\includegraphics[width=\linewidth]{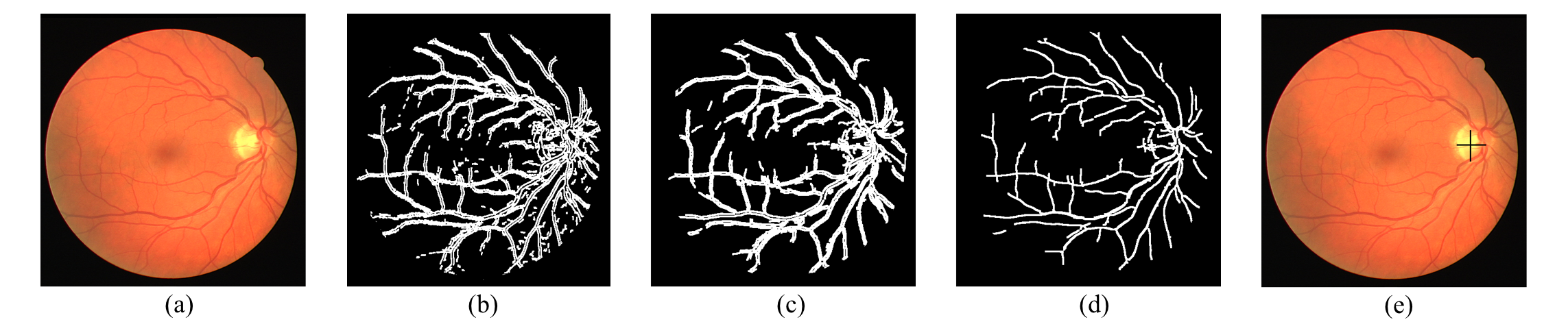}
\centering
\caption{(a) Input image from DRIVE; (b) result of vessel segmentation using Canny; (c) result of vessel segmentation using LoG; (d) result of vessel segmentation using Match filter; (e) final ONH detection.}
\label{fig6}
\end{figure}

\begin{figure}[h!]
\includegraphics[width=\linewidth]{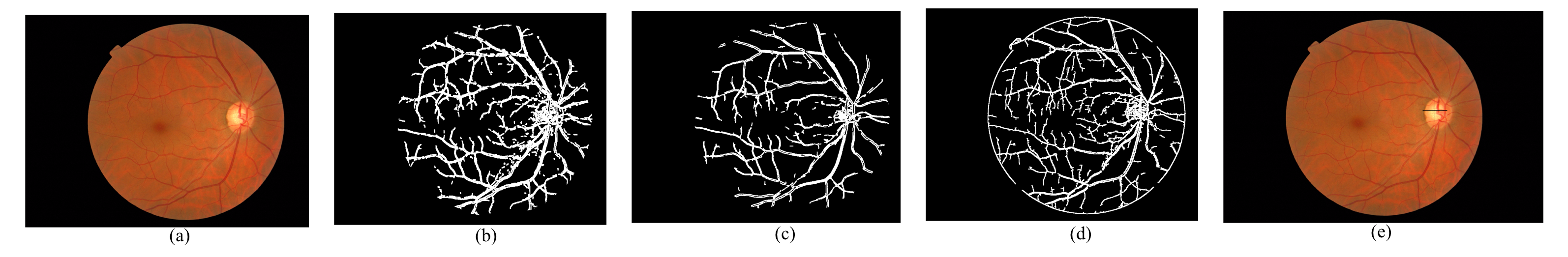}
\centering
\caption{(a) Input image from DRIVE; (b) result of vessel segmentation using Canny; (c) result of vessel segmentation using LoG; (d) result of vessel segmentation using Match filter; (e) final ONH detection.}
\label{fig7}
\end{figure}

\section{Discussion and Conclusion}
A potential use of fundal digital image analysis is the ability to analyze large fundal images in a short period of time without
tiredness. 
The identification of fundal landmark features such as the ONH, fovea and the retinal vessels as reference
coordinates is a prerequisite before systems can do more complex tasks identifying pathological entities.
Reliable techniques exist for identification of these structures in retinal photographs \cite{Khansari-DR, Khansari-TMI, Khansari-optics, tavakoli-fluoresceinangiography, Tavakoli-ONH1}.\\
Since fundus images are nowadays in digital format, it is possible to create a computer-based system that automatically detects
abnormal lesions from fundus images \cite{welikala, welikala2}. 
An automatic screening system would save the time of well-paid clinicians,
letting eye clinics to use their ophthalmologists in other important tasks. 
It could also be possible to screen more people
and more often with the help of an automatic screening system, since it would be more inexpensive than screening by humans.
In this study an automated algorithm were utilized for ONH detection without intervention of any ophthalmologist. 
We evaluate three vessel segmentation methods for detection of ONH.
We presented the segmentation of the ONH by separately using of combination of Canny edge detector, LoG edge detector, and Match filtert and multi-overlapping window. 
The quality of the ONH detection depends on some parameters such as the window size (n), number of step, thresholding validation, etc.
Besides training set of images is completely independent and they  selected randomly. 
 Our automated system has been developed using color retinal images provided by DRIVE and CHASE-DB public databases of color fundus images and one local database MUMS-DB consist of normal and diabetic retinal images. 
DRIVE consists of 40 images, divided into training and a test set, both containing 20 images. The CHASE-DB consist of 28 fundus images from both left and right eyes. And the MUMS-DB consists of 100 fundus images all images are obtained using 45° Non-Mydriatic retinal camera. \\
In the final section, statistical information about the sensitivity and specificity measures is extracted. The higher the sensitivity and specificity values, the better the procedure. For all retinal images of test set (120 images), our reader labeled the ONH on the images and the result of this manual detections were saved to be analyzed further. According to manual ONH detection using the Laplacian-of-Gaussian vessel segmentation our automated algorithm finds 18 of ONH in true location for 20 color images in CHASE-DB database and all images in DRIVE database. For the local database, MUMS-DB, our method detected the ONH correctly in 72 images out of 80 images.
The Canny vessel segmentation our automated algorithm finds 15 of ONH in true location for 20 color images in CHASE-DB database and 16 out of 20 images in DRIVE database. For the local database, MUMS-DB, our method detected the ONH correctly in 70 images out of 80 images.
At last, using matched filter in the vessel segmentation our algorithm finds 19 of ONH in true location for 20 color images in CHASE-DB database and all images in DRIVE database. For the local database, MUMS-DB, our method detected the ONH correctly in 75 images out of 80 images. \\
From sensitivity and specificity view point, our ONH detection algorithm in comparing with our three different vessel segmentation methods we received more than 90\%, and 95\% for both sensitivity and specificity for all color retinal images respectively in three databases. It is better than some reports which were concentrated on ONH detection. One limitation of the current methods is that, when we have a lesion nearly same size and same intesity of the ONH the algorithm take that as the ONH like Fig.~\ref{fig8}. The reason for that probably because of inaccurate vessel segmentation or preprocessing steps. On future work we will work on this rpoblem to solve it.\\
Although we didn't work
on ONH boundary detection and only use a template to mask the ONH but our results is acceptable in covering the ONH more
completely while save other retinal area unmasked.
On other hand, as we said Sinthanayothin et al. \cite{sinthanayothin}, detected ONH but, others have found that their algorithm often fails for
fundus images with a large number of white lesions, light artifacts or strongly visible choroidal vessels \cite{Lowell J}. Others have
exploited the Hough transform (a general technique for identifying the locations and orientations of certain types of shapes
within a digital image; \cite{LKalviainen}, to locate the ONH \cite{Tamura}. However, Hough spaces tend to be sensitive to the chosen image
resolution \cite{hoover-fuzzyconvergence}.
Even the ONH detection is useful for pattern of some diseases such as glaucoma \cite{GoldbaumGlucom, Septiarini}.
The goal of this work was to develop algorithms for detecting different abnormal vascular lesions related to DR.
\\
The results show that all the  three segmentation methods did acceptable results in ONH detection. Among them using Match filter worked better than the others. From sensitivity and specificity view point, our ONH detection algorithm in comparing with our three different vessel segmentation methods we received more than 90\%, and 95\% for both sensitivity and specificity for all color retinal images respectively in three databases. 
Our algorithm also has some important characteristics in the detection of vascular structure in retinal images that include: it’s robustness to noise, acceptable performance in the detection of both thick and thin vessels by the combined methods and multi-overlapping windows, and last but not least, simplicity of the whole method in comparison with other methods mentioned in this paper.


%

\appendices


\ifCLASSOPTIONcaptionsoff
  \newpage
\fi

\end{document}